\documentclass[10pt,epsf]{article}

\usepackage{graphicx}
\usepackage{indentfirst}
\usepackage{cite}
\usepackage{color}
\usepackage{subfigure}

\begin{document}

\title{\bf{The Upper Bound of Radiation Energy \\in the Myers-Perry Black Hole Collision}}

\date{}
\maketitle

\begin{center}
\author{Bogeun Gwak}$^a$\footnote{rasenis@sogang.ac.kr}, \author{Bum-Hoon Lee}$^{a, b, c, d}$\footnote{bhl@sogang.ac.kr}\\
\vskip 0.25in
$^{a}$\it{Center for Quantum Spacetime, Sogang University, Seoul 04107, Republic of Korea}\\
$^{b}$\it{Department of Physics, Sogang University, Seoul 04107, Republic of Korea}\\
$^{c}$\it{Asia Pacific Center for Theoretical Physics, Pohang 37673, Republic of Korea}\\
$^{d}$\it{Department of Physics, Postech, Pohang 37673, Republic of Korea}\end{center}
\vskip 0.6in

{\abstract
We have investigated the upper bound of the radiation energy in the head-on collision of two Myers-Perry black holes. Initially, the two black holes are far away from each other, and they become one black hole after the collision. We have obtained the upper bound of the radiation energy thermodynamically allowed in the process. The upper bound of the radiation energy is obtained in general dimensions. The radiation bound depends on the alignments of rotating axes for a given initial condition due to spin-spin interaction. We have found that the collision may not be occurred for a initially ultra-spinning black hole.
}

\thispagestyle{empty}
\newpage
\setcounter{page}{1}

\section{Introduction}
Black holes play an important role in the strong gravitational field. The merging of two black holes is one possible source of gravitational waves in our universe. Recently, gravitational waves were detected for the first time by the Laser Interferometer Gravitational-Wave Observatory (LIGO)\cite{Abbott:2016blz}. They originate from a binary black hole merging, with energy equivalent to three times of the solar mass. Massive black holes may be important in studying the early universe in the consideration of the Higgs particle discovered at the Large Hadron Collider (LHC)\cite{ATLAS:2012ae,Chatrchyan:2012tx}. In the Higgs potential, the running of the coupling may imply that the present universe is metastable, so that the universe decays into true vacua with certain lifetime. Compared with the age of the universe, the decay lifetime can be long enough due to the large energy barrier \cite{Coleman:1977py,Callan:1977pt,Coleman:1980aw}. However, the energy barrier can be lowered by inhomogeneities reducing the decay lifetime up to millions of Planck times\cite{Burda:2015isa,Burda:2015yfa}. The black hole, a gravitational impurity, can generate such inhomogeneities.

Cosmic censorship conjecture suggests that the black hole horizon should stably cover the singularity\cite{Penrose:1969pc}. The stabilities of black holes are tested to prove the cosmic censorship conjecture after the black hole absorbs external fields or particles. The validity of the conjecture depends on the black hole. The cosmic censorship conjecture is known to be valid in Kerr black holes\cite{Wald:1974ge,Jacobson:2009kt,Barausse:2010ka,Barausse:2011vx,Colleoni:2015afa,Colleoni:2015ena,Cardoso:2015xtj}. Although the conjecture is invalid in near extremal Reissner-Nordstr\"{o}m (RN) black holes\cite{Hubeny:1998ga}, it becomes valid when back-reaction is considered\cite{Isoyama:2011ea}. Cosmic censorship conjecture is valid in tests for Myers-Perry (MP)\cite{Myers:1986un,BouhmadiLopez:2010vc,Doukas:2010be,Saa:2011wq,Gao:2012ca} and anti-de Sitter (AdS)\cite{Rocha:2014gza,Rocha:2014jma,McInnes:2015vga,Rocha:2011wp,Gwak:2015fsa,Gwak:2015sua,Natario:2016bay} black holes, but the violation of the conjecture can occur in black strings and rings\cite{Lehner:2010pn,Figueras:2015hkb}. Therefore, the cosmic censorship still remains as a conjecture.

The black hole is regarded as a thermodynamic system having temperature\cite{Hawking:1974sw,Hawking:1976de}. The energy of the black hole can be extracted through the Penrose process\cite{Bardeen:1970zz,Penrose:1971uk}, and the extracted energy is the rotational energy in the case of Kerr black hole. However, even if the rotational energy can be decreased, the other constituents of energy always increase; this is known as irreducible mass\cite{Christodoulou:1970wf,Christodoulou:1972kt}. The irreducible mass is distributed to the surface of the horizon\cite{Smarr:1972kt}, and the square of the irreducible mass is proportional to the Bekenstein-Hawking entropy\cite{Bekenstein:1973ur,Bekenstein:1974ax}. The laws of thermodynamics are constructed according to the definitions of entropy and temperature in the black hole system. These laws are necessary to valid the cosmic censorship conjecture of Kerr and MP black holes\cite{Gwak:2011rp}.

An MP black hole is the higher-dimensional rotating black hole. The instability of the MP black hole is investigated in higher dimensions. The five-dimensional MP black hole with a single rotation has been found unstable by nonlinear numerical evolution\cite{Shibata:2009ad}, but there have been no unstable modes under the gravitational perturbation\cite{Dias:2014eua}. In over five dimensions, the angular momentum of the MP black hole may have no upper bound. For example, the MP black hole with a single rotation can rapidly rotate. However, angular momentum may have a critical value above which the MP black hole becomes unstable under perturbation\cite{Dias:2009iu,Dias:2010eu,Dias:2010maa}. Instabilities are also found in multi-rotational or extremal cases\cite{Dias:2014eua,Durkee:2010qu,Murata:2012ct}. Non-perturbative fragmentation instability has also been found\cite{Emparan:2003sy}, which makes the black hole at a rapid angular velocity break into two black holes. The fragmentation instability due to thermodynamic preference has been found in AdS black holes\cite{Gwak:2014xra,Gwak:2015ysa} and also in dilatonic Gauss-Bonnet black holes\cite{Ahn:2014fwa}.

Two rotating black holes radiate energy in their coalescing process. An upper bound for the radiation energy has been found using the laws of thermodynamics\cite{Hawking:1971tu}. The radiation energy depends on the alignment of rotating axes of two black holes due to spin-spin interaction\cite{Schiff:1960gi,Wilkins:1970wap,Mashhoon:1971nm}, which is approximately proportional to the angular momentum of the black hole and equivalent to the energy upper bound for radiation\cite{Wald:1972sz}. The spinning black hole binary can have large enough interaction to contribute to the black hole stability\cite{Majar:2012fa,Zilhao:2013nda}, even at higher dimensions\cite{Herdeiro:2008en}. Such interaction has been also studied for spinning particles in a curved spacetime\cite{Plyatsko:2015bia,d'Ambrosi:2015xci}. The upper bound of radiation is lowered in the high-energy collision of the black holes\cite{Eardley:2002re,Sperhake:2008ga,Coelho:2012sya}. The collision between two neutron stars is thermodynamically studied\cite{Hennig:2006ik,Hennig:2007qg}. Detailed numerical analysis of black hole collision has also been carried out. The basic framework was developed to study head-on collisions for the Einstein gravitational field equations\cite{Smarr:1976qy,Smarr:1977fy,Smarr:1977uf,Witek:2010xi}. The waveform of the gravitational radiation has been obtained with consideration of the tidal heating of horizons in the head-on collision of two black holes\cite{Anninos:1993zj,Anninos:1994ay,Anninos:1998wt}. In five and six dimensions, the values of the radiations are obtained for non-spinning black holes\cite{Zilhao:2010sr,Witek:2010az,Witek:2014mha}. Numerical investigations were recently carried out to obtain the gravitational radiation for various initial conditions of black holes\cite{Reisswig:2009us,Bantilan:2014sra,Bednarek:2015dga,Hirotani:2015fxp,Sperhake:2015siy}.

In this paper, the radiation upper bound is investigated for the collision of two MP black holes. We generalize the Kerr black hole case\cite{Hawking:1971tu,Wald:1972sz} and obtain the bound for radiation energy that is thermodynamically allowed in general dimensions, including four dimensions. The bounds are larger than exact values in numerical simulation for four dimensions, so we focus on overall behaviors of bounds for given dimensions. The radiation bound depends on the alignments of rotating planes due to spin-spin interaction. Over five dimensions with a single rotation, the collision of the two MP black holes cannot be occurred, if one of the black holes rotates rapid velocity. Since there exists a double Myers-Perry Black Hole solution\cite{Herdeiro:2008en}, this behavior is interesting. We will interpret this phenomenon according to the instability and spin-spin interaction of the MP black hole. In multi-rotations, the general behaviors of radiation bounds are shown for various initial conditions.

The paper is organized as follows. In section~\ref{sec2}, we introduce the MP black hole with multi-rotations. In section~\ref{sec3}, we introduce the basic framework and assumptions of the analytical method. Using the framework, we obtain the radiation bounds for general dimensions in the numerical method. In section~\ref{sec4}, we briefly summarize our results.

\section{Myers-Perry Black Holes}~\label{sec2}
The MP black hole is the generalization of the Kerr black hole to higher dimensions in Einstein gravity\cite{Myers:1986un}. The black hole solution is asymptotically flat with multi-rotating planes in higher dimensions. The each rotating plane is defined by every two independent spatial dimensions. Including mass parameter $\mu$ and $i$th spin parameter $a_i$, the $D$-dimensional MP black hole metric is
\begin{eqnarray}\label{eq:metric-multi}
&&ds^2=-dt^2+\frac{U}{V-\mu}dr^2+\frac{\mu}{U}\left[dt-\sum_{i=1}^{n-\epsilon}a_i\gamma^2_id\phi_i\right]^2+\sum_{i=1}^{n}(r^2+a_i^2)d\gamma_i^2+\sum_{i=1}^{n-\epsilon}(r^2+a_i^2)\gamma_i^2d\phi_i^2\,\\
&&U=r^{\epsilon}\sum_{i=1}^{n}\frac{\gamma_i^2}{r^2+a_i^2}\prod_{j=1}^{n-\epsilon}(r^2+a_j^2)\,,\,\,F=r^2\sum_{i=1}^{n}\frac{\gamma_i^2}{r^2+a_i^2}\,,\,\,V=r^{\epsilon-2}\prod_{i=1}^{n-\epsilon}(r^2+a_i^2)=\frac{U}{F}\,.\nonumber
\end{eqnarray}
where $n=[D/2]$. The direction cosines $\gamma_i$ are constrained to $\sum_{i=1}^{n}\gamma_i^2=1\,$. The metric of even and odd dimensions is distinguished by an $\epsilon$ value of $1$ for even and $0$ for odd cases. Note that there is no rotation in the last spatial direction for the even-dimension cases. The mass parameter $\mu$ and spin parameters $a_i$ are related to the mass and angular momenta of the black hole as
\begin{eqnarray}
M=\frac{(D-2)\Omega_{D-2}}{16\pi G}\mu\,,\quad J_i=\frac{\Omega_{D-2}}{8\pi G}\mu a_i\,,
\end{eqnarray}
where (D-2)-dimensional solid angle is denoted to $\Omega_{D-2}$. An outer horizon $r_h$ satisfies
\begin{eqnarray}
r_h^{\epsilon-2}\prod_{i=1}^{n-\epsilon}(r_h^2+a_i^2)-\mu=0\,,
\end{eqnarray}
which is rewritten for the MP black hole with a single spin parameter $a$,  
\begin{eqnarray}
r_h^2+a^2-\frac{\mu}{r_h^{D-5}}=0\,.
\end{eqnarray}
Over five dimensions, there is a solution for every values of the spin parameter. So the angular momentum has no Kerr bound in over five dimensions. The single rotating black hole can rotate extremely rapid velocity. The entropy $S_{BH}$ and area $A_H$ of the black hole are
\begin{eqnarray}
S_{BH}=\frac{1}{4}A_H\,,\quad A_H=\Omega_{D-4}\mu r_h\,,
\end{eqnarray}
in which the horizon of the black hole is denoted as $r_h$.

\section{Myers-Perry Black Hole Collisions}~\label{sec3}
We will investigate the upper bounds of radiation energy released after the collision of two MP black hole. We generalize the Kerr black hole case\cite{Hawking:1971tu,Wald:1972sz} to that of the MP black hole. There are two stationary MP black holes far from each other in the initial state. We treat these black holes independently since their interactions are negligible. The first black hole is mass $M_1$ and angular momenta $J_{1i}$. The second black hole is $M_2$ and $J_{2i}$, in which index $i$ signifies the $i$th rotating plane. The rotating planes of black holes are assumed to be aligned parallel or anti-parallel. After the collision, the final state is one MP black hole with mass $M_f$ and angular momenta $J_{fi}$. The collision is a natural process, so the entropy of the final state should be larger than that of the initial state. This is described in terms of the inequality for the sum of the horizon area between the initial and final states,
\begin{eqnarray}
\label{eq:ent01}
A_H(M_1,J_{1i})+A_H(M_2,J_{2i})\leq A_H(M_f,J_{fi})\,.
\end{eqnarray}  
The entropy of radiation is ignored, because the actual radiation is much less than total mass of the initial state. In head-on collision, the angular momentum is conserved since it cannot be radiated away.
\begin{eqnarray}
\label{eq:ang01}
J_{1i}+J_{2i}=J_{fi}\,,\quad i=1,\,\,2,\,...,\,n-\epsilon\,.
\end{eqnarray}
The mass range of the final black hole can be obtained in combination with Eq.~(\ref{eq:ent01}) and (\ref{eq:ang01}). The loss of the mass at the final black hole can be released by the radiation. The radiation energy $M_{rad}$ is
\begin{eqnarray}
\label{eq:radeq02}
M_{rad}=M_1+M_2-M_f\,,
\end{eqnarray}
where the mass $M_f$ is a solution of inequality. So the maximum value of the radiation energy $M_{rad}$ occurs at  the minimum value of the mass $M_f$, which will saturate the inequality in Eq.~(\ref{eq:ent01}). Note that the maximum value of the radiation is the thermodynamically allowed upper bound. We can rewrite the Eq.~(\ref{eq:ent01}) and (\ref{eq:ang01}) 
\begin{eqnarray}
\label{eq:single01a}
\mu_1 r_{1} + \mu_2 r_{2} \leq \mu_f r_{f}\,,\quad \mu_1 a_{1i}+\mu_2 a_{2i}=\mu_f a_{fi}\,,\quad i=1,\,\,2,\,...,\,n-\epsilon\,.
\end{eqnarray}
where the horizons of the black holes are $r_1$, $r_2$, and $r_f$. To describe the radiation energy simply as the same dimensionality of the mass parameter, the radiation energy is redefined as $\mu_{rad}$
\begin{eqnarray}\label{eq:murad5}
\mu_{rad}=\mu_1+\mu_2-\mu_f\,,
\end{eqnarray}
which is rewritten from Eq.~(\ref{eq:radeq02}). Technically, the bound of the radiation energy $\mu_{rad}$ can be obtained by solving $(n-\epsilon+1)$ equations in Eq.~(\ref{eq:single01a}) saturating the inequality.

The radiation energy generally depends on the initial alignment of the black holes. In the approximation for the small and slowly rotating second black hole with a single rotation, the derivative of radiation $\mu_{rad}$ with respect to $a_{2}$ is from Eq.~(\ref{eq:single01a})
\begin{eqnarray}\label{eq:rad03}
\frac{\partial \mu_{rad}}{\partial a_2}=-\frac{2a_1 \mu_1 \mu_2 \mu_f}{(D-4)r_f^{5-D}\mu_f^3+2(\mu_1^2 a_1^2+\mu_f^2 r_f^2)}+\mathcal{O}(a_2)\,.
\end{eqnarray}
When the black holes rotate parallel, the radiation decreases. The radiation increases for anti-parallel rotation. This means that anti-parallel rotating black holes are more stable than parallel rotating cases. In other words, black holes effectively attract each other in anti-parallel rotation, but they repulse each other in parallel rotation. This corresponds to the spin-spin interaction between spinning black holes\,\cite{Wald:1972sz}. However, this is about slowly rotating black hole case. For highly spinning black holes over five dimensions, we will show that the interaction becomes much complicated and repulsive in the analysis for the radiation bounds. This may be related to doubly rotating MP black hole\cite{Herdeiro:2008en} or instability of the black holes\cite{Emparan:2003sy,Gwak:2014xra}.

Many interesting behaviors can be expected from solving Eq.~(\ref{eq:single01a}). The detailed analysis is shown numerically in the following sections. Note that the parameters will be scaled by $\mu_1$ to dimensionless coordinates for the numerical analysis. The scaled mass parameter $\mu$, spin parameter $a_i$, and horizon $r_h$ are  
\begin{eqnarray}\label{eq:dimless}
\tilde{\mu}=\frac{\mu}{\mu_1}\,,\quad \tilde{a}_i=\frac{a_i}{\mu_1^{1/{D-3}}}\,,\quad \tilde{r}_h=\frac{r_h}{\mu_1^{1/{D-3}}}\,.
\end{eqnarray}
Equivalently, the mass parameter of the first black hole can be set to unity. We will omit tildes for simplicity.

\subsection{Radiation Bounds in Four and Five Dimensions}~\label{sec31}
\begin{figure}[h]
\subfigure[{The radiation bounds with respect to $a_2$. The black hole is $\mu_2=1$.}]
{\includegraphics[scale=0.85,keepaspectratio]{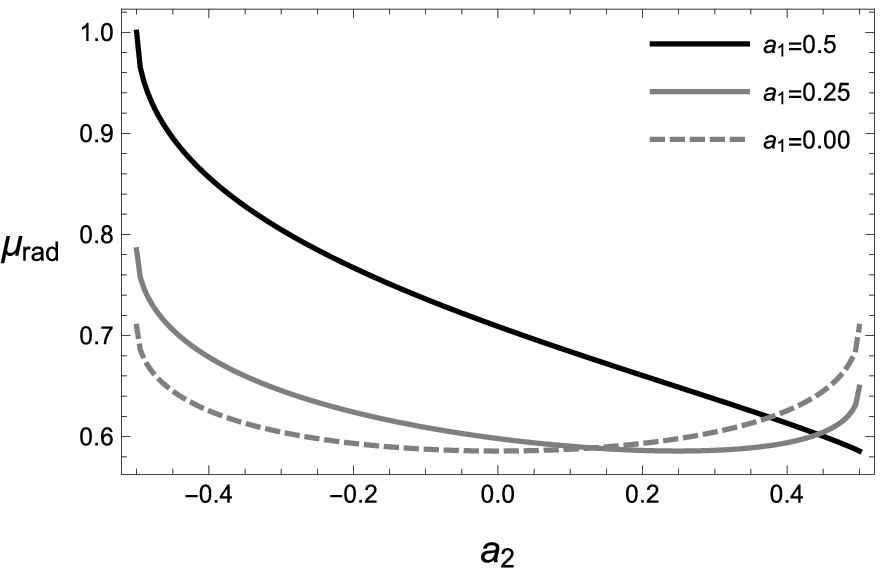}}
\hfill
\subfigure[{The radiation bounds with respect to $\mu_{2}$. The black hole is $a_{1}=0.5$.}]
{\includegraphics[scale=0.85,keepaspectratio]{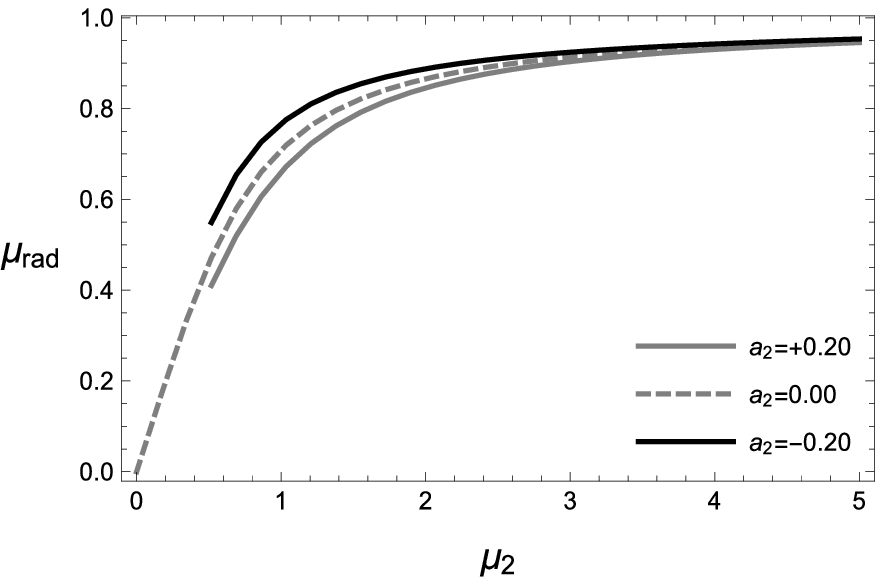}}
\caption{{\small The radiation bounds for 4-dimensional MP black hole collisions.}}
\label{fig1-45D}
\end{figure}

The MP black hole metric in four-dimensional spacetime is that of the Kerr black hole. The radiation bounds are saturated the inequality in Eq.~(\ref{eq:radeq02}), so actual radiation is under the bound\cite{Hawking:1971tu}. Now, we use the dimensionless parameters defined in Eq.~(\ref{eq:dimless}) in which the mass parameter of the first black hole is set to unity. The interaction between Kerr black holes depends on the alignment of rotating axes, which is interpreted as attraction and repulsion\cite{Wald:1972sz}. The radiation bound depends on the alignment of initial black holes. The radiation bounds of anti-parallel cases are larger than those of parallel cases, as shown in Fig.~\ref{fig1-45D}.
The radiation bounds increase when the difference of the rotating velocities become larger, as shown in Fig.~\ref{fig1-45D}~(a). Anti-parallel alignment radiates more energy than is radiated in parallel cases because the radiation slope is negative for spin parameter $a_2$ in Eq.~(\ref{eq:rad03}). The minimum points appear in parallel alignment. The minimum moves to a positive direction with respect to $a_2$ for larger values of $a_1$. The radiation bound increases along with the second black hole mass $\mu_2$, as shown in Fig.~\ref{fig1-45D}~(b).
\begin{figure}[h]
\subfigure[{The radiation bounds with respect to $a_2$. The black hole is $\mu_2=1$.}]
{\includegraphics[scale=0.85,keepaspectratio]{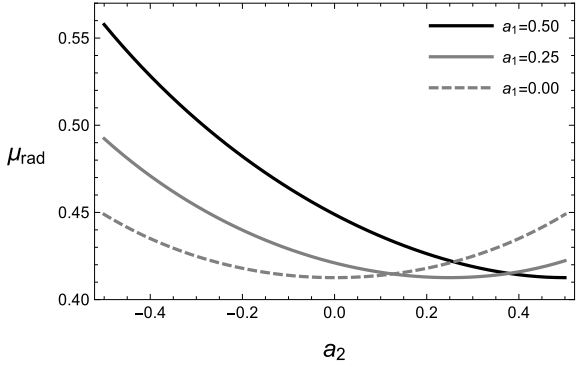}}
\hfill
\subfigure[{The radiation bounds with respect to $\mu_{2}$. The black holes are $a_{11}=0.2$ and $a_{12}=a_{22}=0$.}]
{\includegraphics[scale=0.85,keepaspectratio]{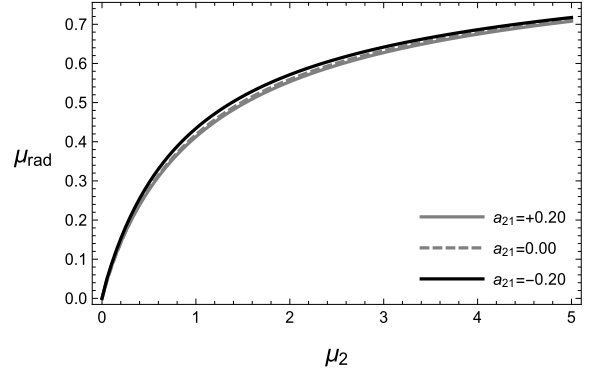}}
\caption{{\small The radiation bounds for the collision of 5-dimensional MP black hole with a single rotation.}}
\label{fig1-5D}
\end{figure}
The bounds for angular momentum restrict the range of the lines. The radiation bounds are still larger in anti-parallel rotations because the interaction energy in attraction is released in the collisions.

The radiation bounds in five-dimensional MP black holes behave similarly to those of four-dimensional Kerr black holes, but two angular momenta are possible in five dimensions because there are four spatial dimensions to be defined in the two rotating planes. In addition, the angular momenta have Kerr bounds. Our analysis is provided within these bounds. As shown in Fig.~\ref{fig1-5D}, The general behaviors of the radiation with respect to $a_2$ and $\mu_2$ are similar to those of four-dimensional cases. Anti-parallel black holes radiate more energy than parallel black holes. The radiation bounds increase along with the mass parameter $\mu_2$ of the second black hole. However, as shown in Fig.~\ref{fig1-5D}~(a), the volumes of the radiation bounds are smaller than those of four-dimensional cases. This change can be also observed in Fig.~\ref{fig1-5D}~(b), in which the largest energy bound is emitted for anti-parallel rotations and the smallest energy in parallel cases. The energy difference between anti-parallel and parallel alignments becomes smaller than that of four-dimensional cases, as shown in Fig.~\ref{fig1-5D}~(b). Compared with four-dimensional MP black holes, the power of the gravitational force increases, so that more energy can be captured by the gravity of the black hole. This is probably the reason for the energy decrease in five-dimensional spacetime.

\begin{figure}[h]
\subfigure[The radiation bounds with respect to $a_{21}$. The black holes are $\mu_2=1$ and $a_{11}=a_{12}=0.2$.]
{\includegraphics[scale=0.85,keepaspectratio]{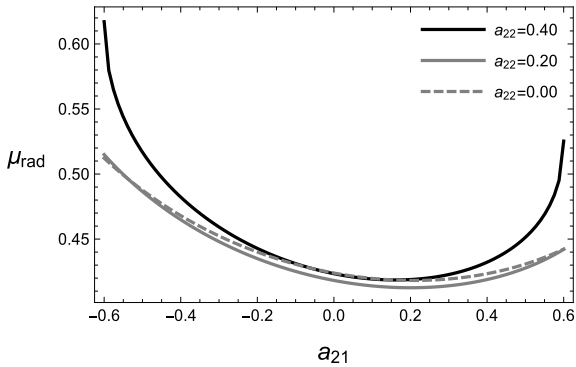}}
\hfill
\subfigure[{The radiation bounds with respect to $a_{12}$. The black holes are $\mu_2=1$ and $a_{11}=a_{22}=0.2$.}]
{\includegraphics[scale=0.85,keepaspectratio]{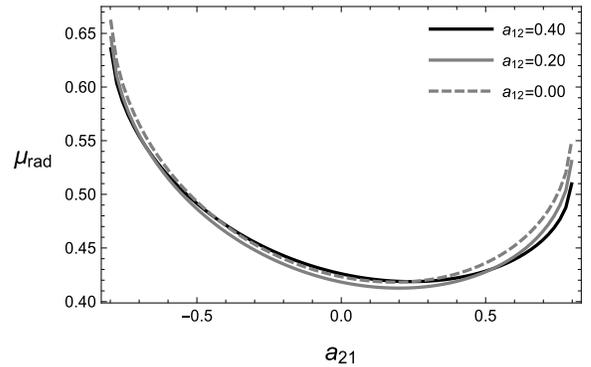}}
\caption{{\small The radiation bounds for the collisions of 5-dimensional MP black hole with two rotations.}}
\label{fig2-5Dtwo}
\end{figure}
Turned on the second angular momenta, $a_{12}$ and $a_{22}$, the radiation bounds are changed in the five-dimensional MP black hole as shown in Fig.~\ref{fig2-5Dtwo}. General shapes for variables, $a_{12}$ and $a_{22}$, are similar to a single rotation in five-dimensional black holes, but these show that the radiation contribution of angular momenta is not linear. Because the positive spin parameters are used, the minimum bound of the radiation in Fig.~\ref{fig2-5Dtwo} appears with these parameters. The radiation increases with large spin parameter $a_{22}$, as shown in Fig.~\ref{fig2-5Dtwo}~(a). However, the effects with respect to spin parameter $a_{12}$ are different from those of $a_{22}$, as shown in Fig.~\ref{fig2-5Dtwo}~(b).

\subsection{Radiation for a Single Rotation over Five Dimensions}~\label{sec32}
\begin{figure}[h]
\subfigure[The radiation bounds with respect to $a_2$. The black hole is $\mu_2=0.5$. The radiation energy is zero at the black points as following figures.]
{\includegraphics[scale=0.85,keepaspectratio]{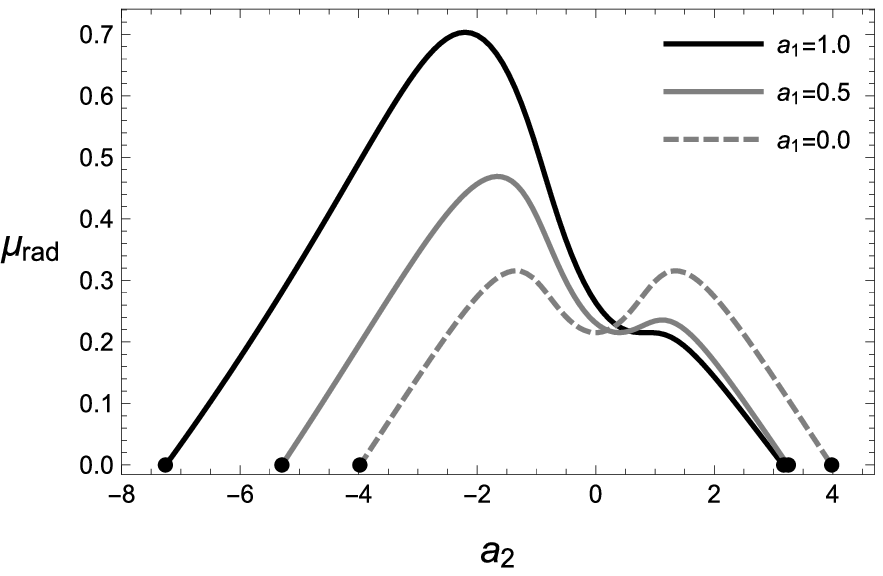}}
\hfill
\subfigure[The radiation bounds with respect to $\mu_2$. The black hole is $a_{2}=4.7$.]
{\includegraphics[scale=0.85,keepaspectratio]{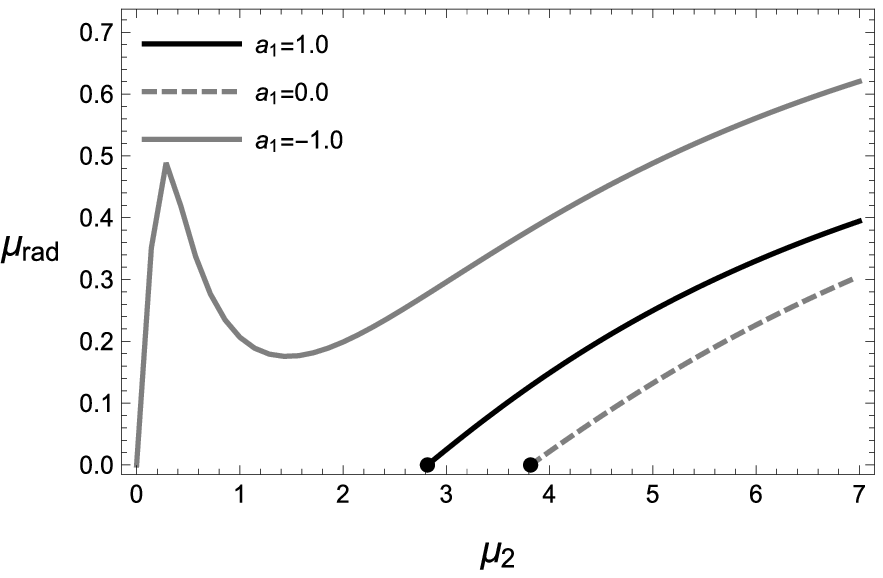}}
\caption{{\small The radiation bounds for the collision of 6-dimensional MP black holes with a single rotation.}}
\label{fig1}
\end{figure}

Over five dimensions, the MP black holes exhibit an explicit difference between single and multiple rotations. For a single rotation, there is no Kerr bound. This makes it possible for the black hole to have an extremely large spin parameter. If one of the initial black holes has large spin parameter with respect to its mass parameter, the final black hole can rapidly rotate. This can make that the entropy of the initial state is larger than that of the final state. Under the assumption that one of the initial black holes is ultra-spinning, $a\gg 1$,
\begin{eqnarray}
r_h\approx \left(\frac{\mu}{a^2}\right)^{\frac{1}{D-5}}\,,\quad \frac{{A}_H(M_f,J_f)}{{A}_H(M_1,J_1)+{A}_H(M_2,J_2)}\approx\left(\frac{1}{a^2}\right)^{\frac{1}{D-5}}\ll 1\,.
\end{eqnarray}
The initial state is thermodynamically preferred, so the collision of two black holes may not be occurred in the higher-dimensional cases. The radiation bounds are numerically obtained as shown in Fig.~\ref{fig1}.
The radiation bounds are shown for 6-dimensional cases with respect to $a_2$ as shown in Fig.~\ref{fig1}~(a). An anti-parallel alignment emits more energy than a parallel case. For fixed $a_2$, the radiation bound increases with the difference of the rotating velocities between the initial black holes. There are two maxima. One appears in anti-parallel alignment, while the other appears in parallel alignment. The maximum in the anti-parallel case is the largest radiation energy because more interaction energy is released in anti-parallel cases than in parallel ones. The minimum is located in parallel alignments between the two maxima. For large spin parameters, the radiation energy decreases to eventually reach zero at the black points as shown in Fig.~\ref{fig1}~(a). For spin parameters beyond black points, the entropy of the initial state is larger than that of the final state. The MP black holes are thermodynamically preferred to stay in the initial states. This is not observed in four and five dimensions. The radiation continuously increases with respect to mass parameter $\mu_2$, as shown in Fig.~\ref{fig1}~(b). The anti-parallel cases emit more energy than parallel cases, as anticipated from Fig.~\ref{fig1-5D}~(b) and \ref{fig2-5Dtwo}~(b). The radiation bounds start at the black points, because the rapid spinning black hole is not thermodynamically preferred to collision. That can be interpreted as demonstrating there are no coalesces in large spin parameters. The phenomena of zero radiation can be related to repulsion in the spin-spin interaction. The interaction is approximately proportional to the spin parameter of the black holes. Therefore, without an angular momentum bound, extremely strong repulsion may be possible for large spin parameter $a_1$ or $a_2$. Different from slowly spinning cases, the interaction between highly spinning black holes might be complicated and repulsive. The strong repulsion with a large spin parameter interrupts the black hole collisions.

\begin{figure}[h]
\subfigure[The radiation bounds with respect to spin parameter $a_2$. The mass parameters are $\mu_2=0.5$.]
{\includegraphics[scale=0.85,keepaspectratio]{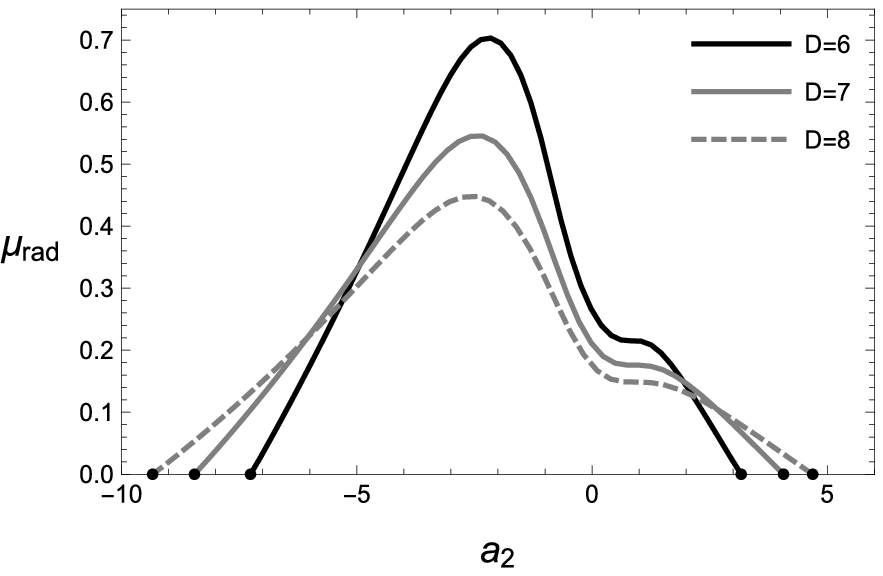}}
\hfill
\subfigure[The graph for values of $a_f/r_f$ for final states. The horizontal lines are values of $a_c/r_h$ for given dimensions. The mass parameters are $\mu_2=0.5$.]
{\includegraphics[scale=0.85,keepaspectratio]{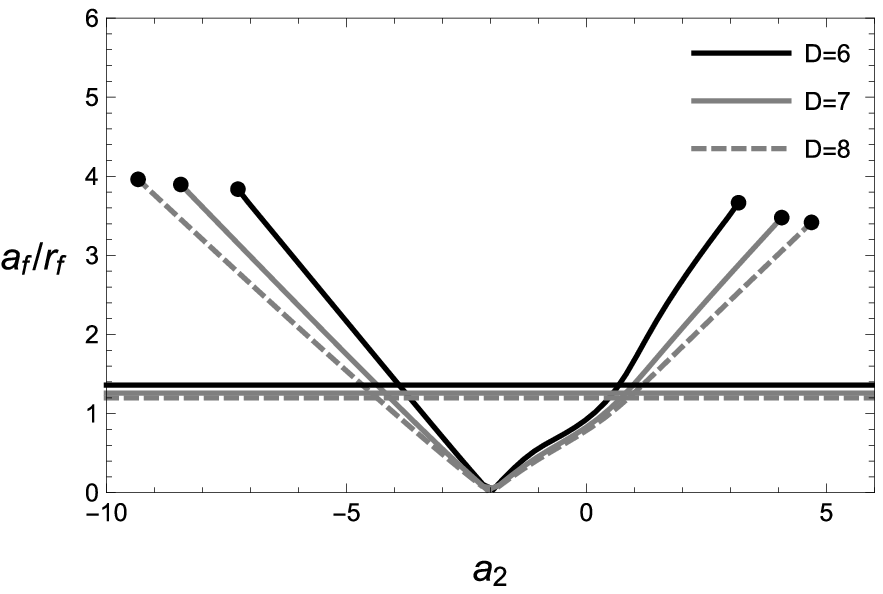}}
\caption{{\small The radiation bounds and critical spin parameters of the collision of higher-dimensional MP black holes with a single rotation.}}
\label{fig2}
\end{figure}

The phenomena of no radiation at the large spin parameter suggest that two black holes may not be merged by the collision. If we assume that the final black hole is formed, then this black hole should be unstable. In other words, the final black hole should be fragmented into two initial black holes. According to fragmentation instability\cite{Emparan:2003sy,Gwak:2014xra}, the MP black hole can be unstable beyond critical spin parameters for given dimensions. Critical spin parameters, $a_c$, are
\begin{eqnarray}
\frac{a_c}{r_h}\geq 1.36\,\,\textrm{in}\,\,D=6\,,\quad\frac{a_c}{r_h}\geq 1.26\,\,\textrm{in}\,\,D=7\,,\quad\frac{a_c}{r_h}\geq 1.20\,\,\textrm{in}\,\,D=8\,,
\end{eqnarray}
The phenomena of no radiation appears for higher-dimensional spacetime, as shown in Fig.~\ref{fig2}~(a).
For the same mass and spin parameters, the radiation bounds have smaller maxima and wider spin parameter ranges in higher dimensions. In addition, the MP black holes can be thermodynamically preferred for the initial states over five dimensions. This can be interpreted that if the black hole of the final state is unstable due to fragmentation instability. To illustrate this interpretation, we test whether the final black hole can undergo fragmentation instability. All spin parameters of no radiation are beyond the critical spin parameters as shown in Fig.~\ref{fig2}~(b). The phenomena of no radiation appear at each black point, and fragmentation instability occurs over each horizontal line. Therefore, the final state can be unstable if one of the initial black holes is rapidly spinning. More energy is needed to form the unstable final black hole, and then the radiation energy decreases to zero.

\subsection{Radiation for Multi-Rotations over Five Dimensions}~\label{sec33}
Multi-rotating MP black holes have angular momentum bounds for each rotating plane over five dimensions. This is different from single-rotation cases.
\begin{figure}[h]
\subfigure[The radiation bound with respect to $a_{21}$ for $\mu_2=0.5$ and $a_{12}=a_{22}=0.01$.]
{\includegraphics[scale=0.6,keepaspectratio]{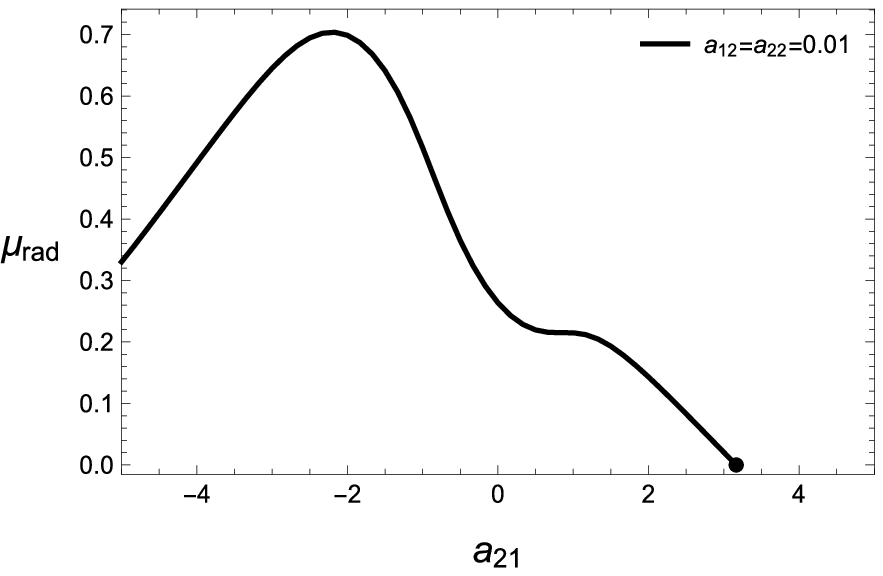}}
\hfill
\subfigure[The radiation bound with respect to $a_{21}$ for $\mu_2=0.5$ and $a_{12}=a_{22}=0.05$.]
{\includegraphics[scale=0.6,keepaspectratio]{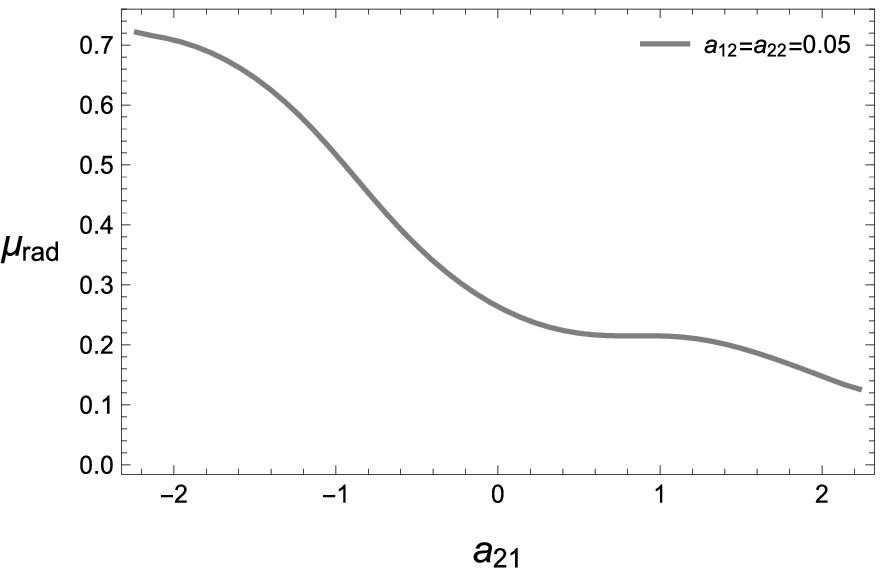}}
\hfill
\subfigure[The radiation bound with respect to $a_{21}$ for $\mu_2=0.5$ and $a_{12}=a_{22}=0.2$.]
{\includegraphics[scale=0.6,keepaspectratio]{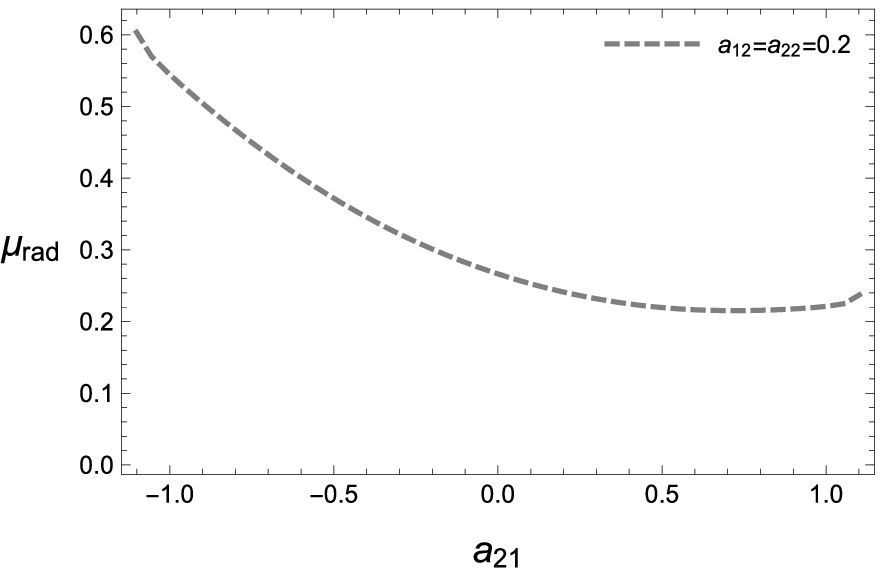}}
\caption{{\small The radiation bounds for the collision of 6-dimensional MP black holes with two rotations.}}
\label{fig6multi}
\end{figure}
Due to Kerr bounds for angular momenta, the ranges of the radiations are also bounded. At wide angular bounds, the radiation bounds are similar to those of a single rotation, but the shapes of the bounds become similar to those of five-dimensional cases, as shown in Fig.~\ref{fig6multi}.
The second rotating plane is turned on in Fig.~\ref{fig6multi}~(a) from Fig.~\ref{fig1}~(a). There is still a zero radiation point in positive $a_{21}$ because the spin-spin repulsion is stronger in parallel cases. For the smaller range, the zero radiation range disappears in Fig.~\ref{fig6multi}~(b).
\begin{figure}[h]
\subfigure[The radiation bounds with respect to $a_{21}$ in 6-dimensional MP black holes. The parameters are $\mu_2=0.5$, $a_{11}=1$, and $a_{12}=a_{22}=0.2$.]
{\includegraphics[scale=0.85,keepaspectratio]{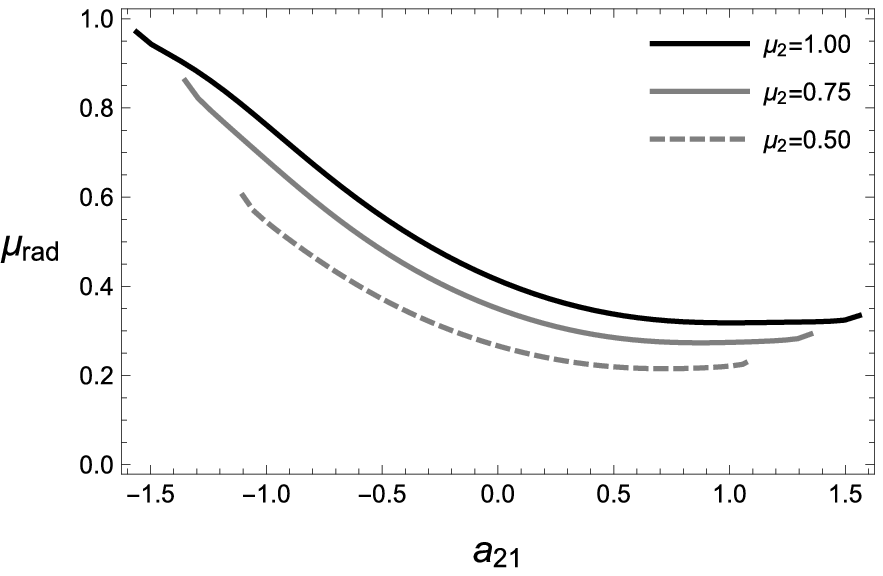}}
\hfill
\subfigure[The radiation bounds with respect to $a_{21}$. The parameters are $\mu_2=0.5$, $a_{11}=1$, and $a_{12}=a_{22}=0.2$.]
{\includegraphics[scale=0.85,keepaspectratio]{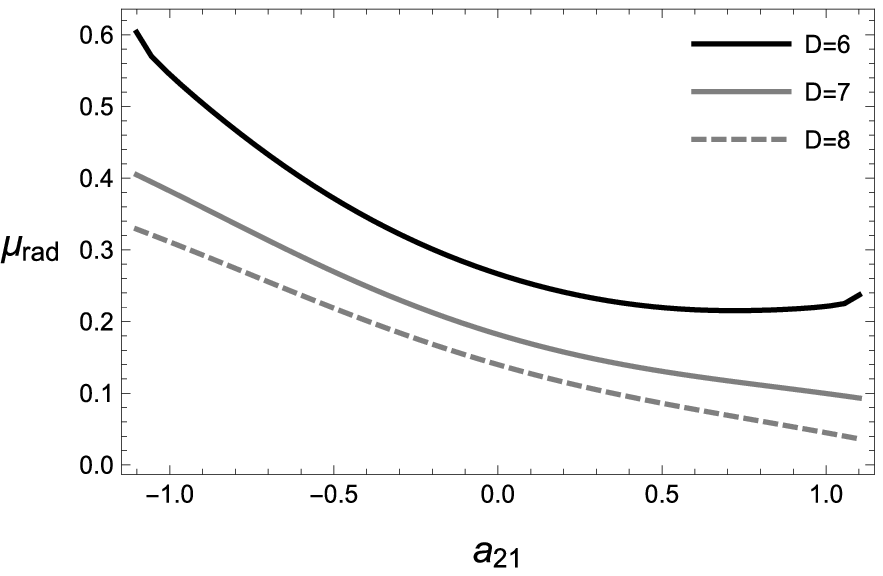}}
\caption{{\small The radiation bounds for the collision of higher-dimensional MP black holes with two rotations.}}
\label{fig7}
\end{figure}
Finally, the radiation bound in Fig.~\ref{fig6multi}~(c) becomes similar to those of five-dimensional cases. This behavior is commonly observed over five-dimensional spacetimes with multi-rotations. Note that the effect of turned on spin parameters $a_{12}$ and $a_{22}$ is very small, so the lines in Fig.~\ref{fig6multi} almost overlap. The radiation bounds increase along with the mass parameter $\mu_2$, as shown in Fig.~\ref{fig7}~(a). In the larger mass, the spin parameter ranges are wider, so the lines are longer for large mass parameters. The radiation strengths decrease for higher dimensions based on thermodynamics, as shown in Fig.~\ref{fig7}~(b). Note that the radiation energy differently depends on the dimensionality in numerical simulation\cite{Zilhao:2010sr,Witek:2010az,Witek:2014mha}.

\section{Summary and Conclusion}\label{sec4}
We have investigated the radiation bound in the collision of two MP black holes. We assume that in the initial state, the two stationary black holes are far away from each other, and then they collide head-on. In the final state, they will form one MP black hole with some energy loss by radiation. In this process, we have obtained the thermodynamic radiation bounds by comparing the entropies between initial and final states.

The radiations in anti-parallel alignment are expected to be larger than those in parallel cases as can be seen in Eq.~(\ref{eq:rad03}). That is related to the spin-spin interaction, attraction at anti-parallel and repulsion at parallel alignment. There exists some qualitative difference depending on whether the spacetime dimension is larger than five or not. In four and five dimensions, the radiation bounds become larger in anti-parallel alignments, and the radiation minimum appeared in parallel cases. This result comes from the release of the interaction energy. The radiation energy is also proportional to the mass parameters of the black holes. 

In cases with more than five dimensions and a single rotation, there are two maxima in the radiation energy with respect to the spin parameter. The overall radiations are large in the anti-parallel alignment. However, for large spin parameters, the radiation bounds decrease to zero. This implies that the initial state is thermodynamically preferred rather than the collision. This result comes from the lack of bounds for angular momentum of the black holes. The spin-spin interaction becomes more complicated in high spinning cases. The interaction becomes repulsion and increases with spinning velocity, so the repulsion is expected to be too strong for the two black holes to collide. Otherwise, this behavior can be interpreted as the instability of the black hole. We considered fragmentation instability. If one of the initial black holes has large spin parameter, the final black hole has been unstable under the fragmentation. Therefore, the formation of the unstable final black hole deceases the radiation energy to zero.

In cases with more than five dimensions and a multi-rotation, the bounds of angular momenta are evident. The radiation behaviors are similar those in single-rotation cases in small second spin parameters, and the behaviors become identical to those of four and five dimension cases with large second spin parameters.
\\

{\bf{Acknowledgments}}

{\small This research was supported by Basic Science Research Program through the National Research Foundation of Korea(NRF) funded by the Ministry of  Science, ICT \& Future Planning(2015R1C1A1A02037523) and the National Research Foundation of Korea(NRF) grant funded by the Korea government(MSIP)(No.~2014R1A2A1A01002306).}


\begin{thebibliography}{99}

\bibitem{Abbott:2016blz} 
  B.~P.~Abbott {\it et al.} [LIGO Scientific and Virgo Collaborations],
  Phys.\ Rev.\ Lett.\  {\bf 116}, no. 6, 061102 (2016).



\bibitem{ATLAS:2012ae}
  G.~Aad {\it et al.} [ATLAS Collaboration],
  Phys.\ Lett.\ B {\bf 710} (2012) 49.

\bibitem{Chatrchyan:2012tx} 
  S.~Chatrchyan {\it et al.} [CMS Collaboration],
  Phys.\ Lett.\ B {\bf 710}, 26 (2012).

\bibitem{Coleman:1977py} 
  S.~R.~Coleman,
  Phys.\ Rev.\ D {\bf 15}, 2929 (1977).

\bibitem{Callan:1977pt} 
  C.~G.~Callan, Jr. and S.~R.~Coleman,
  Phys.\ Rev.\ D {\bf 16}, 1762 (1977).

\bibitem{Coleman:1980aw} 
  S.~R.~Coleman and F.~De Luccia,
  Phys.\ Rev.\ D {\bf 21}, 3305 (1980).

\bibitem{Burda:2015isa} 
  P.~Burda, R.~Gregory and I.~G.~Moss,
  Phys.\ Rev.\ Lett.\  {\bf 115}, 071303 (2015).

\bibitem{Burda:2015yfa} 
  P.~Burda, R.~Gregory and I.~Moss,
  JHEP {\bf 1508}, 114 (2015).

\bibitem{Penrose:1969pc} 
  R.~Penrose,
  Riv.\ Nuovo Cim.\  {\bf 1}, 252 (1969)
  [Gen.\ Rel.\ Grav.\  {\bf 34}, 1141 (2002)].


\bibitem{Wald:1974ge} 
  R.~M.~Wald,
  Annals Phys.\  {\bf 82}, 548 (1974).

\bibitem{Jacobson:2009kt} 
  T.~Jacobson and T.~P.~Sotiriou,
  Phys.\ Rev.\ Lett.\  {\bf 103}, 141101 (2009).

\bibitem{Barausse:2010ka} 
  E.~Barausse, V.~Cardoso and G.~Khanna,
  Phys.\ Rev.\ Lett.\  {\bf 105}, 261102 (2010).

\bibitem{Barausse:2011vx} 
  E.~Barausse, V.~Cardoso and G.~Khanna,
  Phys.\ Rev.\ D {\bf 84}, 104006 (2011).

\bibitem{Colleoni:2015afa} 
  M.~Colleoni and L.~Barack,
  Phys.\ Rev.\ D {\bf 91}, 104024 (2015).

\bibitem{Colleoni:2015ena} 
  M.~Colleoni, L.~Barack, A.~G.~Shah and M.~van de Meent,
  Phys.\ Rev.\ D {\bf 92}, no. 8, 084044 (2015).

\bibitem{Cardoso:2015xtj} 
  V.~Cardoso and L.~Queimada,
  Gen.\ Rel.\ Grav.\  {\bf 47}, no. 12, 150 (2015).

\bibitem{Hubeny:1998ga} 
  V.~E.~Hubeny,
  Phys.\ Rev.\ D {\bf 59}, 064013 (1999).

\bibitem{Isoyama:2011ea} 
  S.~Isoyama, N.~Sago and T.~Tanaka,
  Phys.\ Rev.\ D {\bf 84}, 124024 (2011).

\bibitem{Myers:1986un} 
  R.~C.~Myers and M.~J.~Perry,
  Annals Phys.\  {\bf 172}, 304 (1986).

\bibitem{BouhmadiLopez:2010vc} 
  M.~Bouhmadi-Lopez, V.~Cardoso, A.~Nerozzi and J.~V.~Rocha,
  Phys.\ Rev.\ D {\bf 81}, 084051 (2010).

\bibitem{Doukas:2010be} 
  J.~Doukas,
  Phys.\ Rev.\ D {\bf 84}, 064046 (2011).

\bibitem{Saa:2011wq} 
  A.~Saa and R.~Santarelli,
  Phys.\ Rev.\ D {\bf 84}, 027501 (2011);

\bibitem{Gao:2012ca} 
  S.~Gao and Y.~Zhang,
  Phys.\ Rev.\ D {\bf 87}, no. 4, 044028 (2013).

\bibitem{Rocha:2014gza} 
  J.~V.~Rocha, R.~Santarelli and T.~Delsate,
  Phys.\ Rev.\ D {\bf 89}, no. 10, 104006 (2014).

\bibitem{Rocha:2014jma} 
  J.~V.~Rocha and R.~Santarelli,
  Phys.\ Rev.\ D {\bf 89}, no. 6, 064065 (2014);

\bibitem{McInnes:2015vga}
  B.~McInnes and Y.~C.~Ong,
  JCAP {\bf 1511}, no. 11, 004 (2015).

\bibitem{Rocha:2011wp} 
  J.~V.~Rocha and V.~Cardoso,
  Phys.\ Rev.\ D {\bf 83}, 104037 (2011).

\bibitem{Gwak:2015fsa} 
  B.~Gwak and B.-H.~Lee,
  JCAP {\bf 1602}, 015 (2016).

\bibitem{Gwak:2015sua} 
  B.~Gwak and B.-H.~Lee,
  Phys.\ Lett.\ B {\bf 755}, 324 (2016).

\bibitem{Natario:2016bay} 
  J.~Natario, L.~Queimada and R.~Vicente,
  arXiv:1601.06809 [gr-qc].

\bibitem{Lehner:2010pn} 
  L.~Lehner and F.~Pretorius,
  Phys.\ Rev.\ Lett.\  {\bf 105}, 101102 (2010).

\bibitem{Figueras:2015hkb} 
  P.~Figueras, M.~Kunesch and S.~Tunyasuvunakool,
  Phys.\ Rev.\ Lett.\  {\bf 116}, no. 7, 071102 (2016).

\bibitem{Hawking:1974sw} 
  S.~W.~Hawking,
  Commun.\ Math.\ Phys.\  {\bf 43}, 199 (1975).

\bibitem{Hawking:1976de} 
  S.~W.~Hawking,
  Phys.\ Rev.\ D {\bf 13}, 191 (1976).

\bibitem{Bardeen:1970zz} 
  J.~M.~Bardeen,
  Nature {\bf 226}, 64 (1970).

\bibitem{Penrose:1971uk} 
  R.~Penrose and R.~M.~Floyd,
  Nature {\bf 229}, 177 (1971).

\bibitem{Christodoulou:1970wf} 
  D.~Christodoulou,
  Phys.\ Rev.\ Lett.\  {\bf 25}, 1596 (1970).

\bibitem{Christodoulou:1972kt} 
  D.~Christodoulou and R.~Ruffini,
  Phys.\ Rev.\ D {\bf 4}, 3552 (1971).

\bibitem{Smarr:1972kt} 
  L.~Smarr,
  Phys.\ Rev.\ Lett.\  {\bf 30}, 71 (1973).

\bibitem{Bekenstein:1973ur} 
  J.~D.~Bekenstein,
  Phys.\ Rev.\ D {\bf 7}, 2333 (1973).

\bibitem{Bekenstein:1974ax} 
  J.~D.~Bekenstein,
  Phys.\ Rev.\ D {\bf 9}, 3292 (1974).

\bibitem{Gwak:2011rp} 
  B.~Gwak and B.-H.~Lee,
  Phys.\ Rev.\ D {\bf 84}, 084049 (2011).

\bibitem{Shibata:2009ad} 
  M.~Shibata and H.~Yoshino,
  Phys.\ Rev.\ D {\bf 81}, 021501 (2010).

\bibitem{Dias:2014eua} 
  O.~J.~C.~Dias, G.~S.~Hartnett and J.~E.~Santos,
  Class.\ Quant.\ Grav.\  {\bf 31}, no. 24, 245011 (2014).

\bibitem{Dias:2009iu} 
  O.~J.~C.~Dias, P.~Figueras, R.~Monteiro, J.~E.~Santos and R.~Emparan,
  Phys.\ Rev.\ D {\bf 80}, 111701 (2009).

\bibitem{Dias:2010eu} 
  O.~J.~C.~Dias, P.~Figueras, R.~Monteiro, H.~S.~Reall and J.~E.~Santos,
  JHEP {\bf 1005}, 076 (2010).

\bibitem{Dias:2010maa}
  O.~J.~C.~Dias, P.~Figueras, R.~Monteiro and J.~E.~Santos,
  Phys.\ Rev.\ D {\bf 82} 104025 (2010).

\bibitem{Durkee:2010qu} 
  M.~Durkee and H.~S.~Reall,
  Class.\ Quant.\ Grav.\  {\bf 28}, 035011 (2011).

\bibitem{Murata:2012ct} 
  K.~Murata,
  Class.\ Quant.\ Grav.\  {\bf 30}, 075002 (2013).

\bibitem{Emparan:2003sy} 
  R.~Emparan and R.~C.~Myers,
  JHEP {\bf 0309}, 025 (2003).

\bibitem{Gwak:2014xra} 
  B.~Gwak and B.-H.~Lee,
  Phys.\ Rev.\ D {\bf 91}, no. 6, 064020 (2015).

\bibitem{Gwak:2015ysa} 
  B.~Gwak, B.-H.~Lee and D.~Ro,
  arXiv:1509.03493 [gr-qc].

\bibitem{Ahn:2014fwa} 
  W.~K.~Ahn, B.~Gwak, B.-H.~Lee and W.~Lee,
  Eur.\ Phys.\ J.\ C {\bf 75}, no. 8, 372 (2015).


\bibitem{Hawking:1971tu} 
  S.~W.~Hawking,
  Phys.\ Rev.\ Lett.\  {\bf 26}, 1344 (1971).

\bibitem{Schiff:1960gi}
  L.~I.~Schiff,
  Proc.\ Nat.\ Acad.\ Sci.\  {\bf 46} (1960) 871.

\bibitem{Mashhoon:1971nm} 
  B.~Mashhoon,
  J.\ Math.\ Phys.\  {\bf 12}, 1075 (1971).

\bibitem{Wilkins:1970wap}
  D.~C.~Wilkins,
  Ann.\ Phys.\  {\bf 61}, 277 (1970).

\bibitem{Wald:1972sz} 
  R.~M.~Wald,
  Phys.\ Rev.\ D {\bf 6}, 406 (1972).

\bibitem{Majar:2012fa} 
  J.~Majar and B.~Mikoczi,
  Phys.\ Rev.\ D {\bf 86}, 064028 (2012).

\bibitem{Zilhao:2013nda} 
  M.~Zilhao, V.~Cardoso, C.~Herdeiro, L.~Lehner and U.~Sperhake,
  Phys.\ Rev.\ D {\bf 89}, no. 4, 044008 (2014).

\bibitem{Herdeiro:2008en} 
  C.~A.~R.~Herdeiro, C.~Rebelo, M.~Zilhao and M.~S.~Costa,
  JHEP {\bf 0807}, 009 (2008).

\bibitem{Plyatsko:2015bia} 
  R.~Plyatsko and M.~Fenyk,
  Phys.\ Rev.\ D {\bf 91}, no. 6, 064033 (2015).

\bibitem{d'Ambrosi:2015xci} 
  G.~d¡¯Ambrosi, S.~Satish Kumar, J.~van de Vis and J.~W.~van Holten,
  Phys.\ Rev.\ D {\bf 93}, no. 4, 044051 (2016).

\bibitem{Eardley:2002re} 
  D.~M.~Eardley and S.~B.~Giddings,
  Phys.\ Rev.\ D {\bf 66}, 044011 (2002).

\bibitem{Sperhake:2008ga} 
  U.~Sperhake, V.~Cardoso, F.~Pretorius, E.~Berti and J.~A.~Gonzalez,
  Phys.\ Rev.\ Lett.\  {\bf 101}, 161101 (2008).

\bibitem{Coelho:2012sya} 
  F.~S.~Coelho, C.~Herdeiro and M.~O.~P.~Sampaio,
  Phys.\ Rev.\ Lett.\  {\bf 108}, 181102 (2012).


\bibitem{Hennig:2006ik} 
  J.~Hennig and G.~Neugebauer,
  Phys.\ Rev.\ D {\bf 74}, 064025 (2006).

\bibitem{Hennig:2007qg} 
  J.~Hennig, G.~Neugebauer and M.~Ansorg,
  Astrophys.\ J.\  {\bf 663}, 450 (2007).



\bibitem{Smarr:1976qy} 
  L.~Smarr, A.~Cadez, B.~S.~DeWitt and K.~Eppley,
  Phys.\ Rev.\ D {\bf 14}, 2443 (1976).

\bibitem{Smarr:1977fy} 
  L.~Smarr,
  Phys.\ Rev.\ D {\bf 15}, 2069 (1977).

\bibitem{Smarr:1977uf} 
  L.~Smarr and J.~W.~York, Jr.,
  Phys.\ Rev.\ D {\bf 17}, 2529 (1978).

\bibitem{Witek:2010xi} 
  H.~Witek, M.~Zilhao, L.~Gualtieri, V.~Cardoso, C.~Herdeiro, A.~Nerozzi and U.~Sperhake,
  Phys.\ Rev.\ D {\bf 82}, 104014 (2010).


\bibitem{Anninos:1993zj} 
  P.~Anninos, D.~Hobill, E.~Seidel, L.~Smarr and W.~M.~Suen,
  Phys.\ Rev.\ Lett.\  {\bf 71}, 2851 (1993).

\bibitem{Anninos:1994ay} 
  P.~Anninos {\it et al.},
  Phys.\ Rev.\ Lett.\  {\bf 74}, 630 (1995).

\bibitem{Anninos:1998wt} 
  P.~Anninos and S.~Brandt,
  Phys.\ Rev.\ Lett.\  {\bf 81}, 508 (1998).


\bibitem{Zilhao:2010sr} 
  M.~Zilhao, H.~Witek, U.~Sperhake, V.~Cardoso, L.~Gualtieri, C.~Herdeiro and A.~Nerozzi,
  Phys.\ Rev.\ D {\bf 81}, 084052 (2010).

\bibitem{Witek:2010az} 
  H.~Witek, V.~Cardoso, L.~Gualtieri, C.~Herdeiro, U.~Sperhake and M.~Zilhao,
  Phys.\ Rev.\ D {\bf 83}, 044017 (2011).

\bibitem{Witek:2014mha} 
  H.~Witek, H.~Okawa, V.~Cardoso, L.~Gualtieri, C.~Herdeiro, M.~Shibata, U.~Sperhake and M.~Zilhao,
  Phys.\ Rev.\ D {\bf 90}, no. 8, 084014 (2014).


\bibitem{Reisswig:2009us} 
  C.~Reisswig, N.~T.~Bishop, D.~Pollney and B.~Szilagyi,
  Phys.\ Rev.\ Lett.\  {\bf 103}, 221101 (2009).

\bibitem{Bantilan:2014sra} 
  H.~Bantilan and P.~Romatschke,
  Phys.\ Rev.\ Lett.\  {\bf 114}, no. 8, 081601 (2015).

\bibitem{Bednarek:2015dga} 
  W.~Bednarek and P.~Banasinski,
  Astrophys.\ J.\  {\bf 807}, no. 2, 168 (2015).

\bibitem{Hirotani:2015fxp} 
  K.~Hirotani and H.~Y.~Pu,
  Astrophys.\ J.\  {\bf 818}, no. 1, 50 (2016).

\bibitem{Sperhake:2015siy} 
  U.~Sperhake, E.~Berti, V.~Cardoso and F.~Pretorius,
  Phys.\ Rev.\ D {\bf 93}, no. 4, 044012 (2016).



\end{thebibliography}
\end{document}